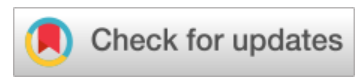

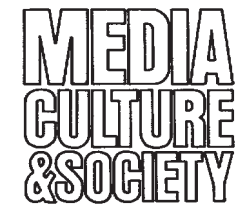

*Main article*

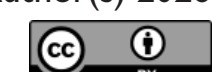

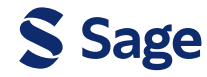

# AI-generated podcasts: Synthetic intimacy and cultural mistranslation in audio overviews from Google's NotebookLM

**Jill Walker Rettberg**[1]

## Abstract

This paper analyses AI-generated podcasts produced by Google's NotebookLM, which generates audio podcasts where two chatty AI hosts discuss whichever documents a user uploads. While AI-generated podcasts have been discussed as tools, for instance in medical education, they have not yet been analysed as media. By uploading different types of text and analysing the generated outputs I show how the podcasts' structure is built around a fixed template. I also find that NotebookLM not only translates texts from other languages into a perky, standardised Mid-Western American English, it also translates cultural contexts to a white, educated, middle-class American default. This is a distinct development in how publics are shaped by media, marking a departure from the multiple public spheres that scholars have described in human podcasting from the early 2000s until today, where hosts spoke to specific communities and responded to listener comments, to an abstraction of the podcast genre. This paper contributes both to research on the use of generative AI in media and to research on cultural bias in LLMs.

[1]University of Bergen, Norway

**Corresponding author:**
Jill Walker Rettberg, Center for Digital Narrative, University of Bergen, Bergen 5020, Norway.
Email: jill.walker.rettberg@uib.no

## Introduction

AI-generated podcasts are the ultimate niche media product. They are generated for an audience of one based on documents uploaded by the user, yet these AI-generated podcasts are also extremely generic. They all have the same style: upbeat, chatty conversations between two American-accented AI hosts that are excited about whatever topic you give them, and encourage you, the listener, to keep asking questions.

AI-generated podcasts, or "Audio Overviews," were first introduced by Google in September 2024, as part of their AI-powered research tool NotebookLM (LM stands for Language Model). While Google pitches its podcast-generating service as a research aid for the individual user, NotebookLM and other services are also used to generate podcasts at scale that are published alongside human podcasts to make money. Individuals began doing this almost as soon as NotebookLM launched the service, and a year later a company like Inception Point AI was producing 3000 episodes a week for its 5000 podcasts. Their CEO told the *Hollywood Reporter* that each episode costs them one US dollar to make, including placing advertisements in it. If 20 people listen to an episode, the company makes a profit.[1] YouTube is rolling out genAI features that will include AI-generated podcasts, and they will generate video for existing podcasts. Mainstream media platforms are using AI tools to to augment their programming. At the same time, we see traces of a backlash: for example, the Washington Post marks their audio versions of articles as "Human Read." AI-generated podcasts are a growing field of media, but media scholars have not yet analysed AI-generated podcasts as media.

In this paper I analyse a range of podcasts generated by Google's NotebookLM to investigate the characteristics of this abstracted genre. I find that although NotebookLM appears to create a customised podcast just for you, it is in fact applying a very particular template to all the podcasts. NotebookLM translates (and mistranslates) sources into Standard American English both linguistically and culturally. I argue that NotebookLM's abstracted model podcast genre is a holdover from the 20th century idea of a shared public sphere, in contrast to the multiple public spheres that are nurtured by human podcasts. Like many human podcasts, the generated podcasts often discuss extremely niche topics, because they discuss whatever sources you give them. However, the generated podcasts discuss niche topics in a universalising way. They use markers of connection that echo the techniques human podcasters use to establish connection with their audiences. But in the AI-generated podcasts, these markers of connection become empty signifiers: words, phrases and vocalisations that lack the actual situatedness of human podcasts. I call this synthetic intimacy.

To develop my argument and identify what is specific to the generated podcast's genre I use a method Gabriele de Seta calls *synthetic probes* (De Seta, 2024). My probes are specific documents that I asked NotebookLM to generate podcasts from, chosen to "stimulate the model in providing data or information about itself" (De Seta, 2024: 18) and listed in Table 1. I chose to probe the model with content that was from markedly different cultures than the educated, middle-class American default that the automated voices and style of the podcast hosts represent. How would the LLM (Large Language Model) translate or refactor material from another country, from another community or from another era? To test this, I first generated podcasts from four selected texts: Norwegian language papers from a faculty board meeting at the University of Bergen, a Norwegian joke, a blog post written in African American Vernacular English, and a 19th century chemistry textbook written as a conversation. Secondly, I generated a set of 20 podcasts from a PDF that consisted of nothing but a single white page. The source documents, the generated audio podcasts and transcripts of the generated podcasts are available as a dataset published on Dataverse (Rettberg, 2026). The examples I have analysed were all generated in the Deep Dive mode, which was the only option for the first months after NotebookLM released the podcast feature, and which is still the default mode.

**Table 1.** Overview of the uploaded documents and how they were translated by NotebookLLM.

| Uploaded source document | Cultural context of source | Language of source | Lost in translation | Added discourse from model |
| --- | --- | --- | --- | --- |
| Meeting documents and agenda for the September 2024 meeting of the Humanities Faculty Board at the University of Bergen | Norwegian academia | Norwegian | Publicly funded education | Budget model of US universities (student tuition)<br>Background about the "crisis in the humanities" |
| Joke | Norwegian | Norwegian | The pun (double meaning of *mus*) | Fairy tales |
| Blog post from Wordonastreet.com | African American culture, hip hop | African American Vernacular English (AAVE) | Rich linguistic play | Polite society |
| Jane Marcet's book *Conversations on Chemistry* (1817) | Early 19th century upper class Britain | English (archaic style) | The conversation and characters in the original, and the careful pedagogy. | Excitement, comparison to contemporary science |
| An empty PDF | None | None | The fact that the document is empty | Conspiracy, puzzle, art |

## Podcasts and multiple public spheres

Traditional broadcast radio is often discussed in terms of the shared public sphere that fosters rational democratic debate in the Habermasian tradition. When podcasts became popular in the early 2000s, they challenged this notion of a shared public sphere, because suddenly anybody could "make radio" without money or governmental permission. This enabled less established or powerful communities to build multiple public spheres.

Jürgen Habermas saw "the borgeouis public sphere" as mediating between the private sphere of the family and trade and the "public" state with its politics. In the 20th century, public broadcasting in many countries was intended to educate the people (Finlayson and Rees, 2023; Habermas, 1991). Broadcast media required expensive equipment and a licence to use the radio frequencies that the signal was broadcast on. Until the 1980s, many countries had public service broadcasters that either had a monopoly or had far greater reach than commercial or volunteer-run radio stations, which often only broadcast locally (Hendy, 2013: 90). Even in countries with fewer restrictions on commercial or volunteer radio broadcasting there was a limited number of radio stations, so a large proportion of the public would listen to the same radio programmes. This allowed society to maintain the idea of a shared public sphere across a nation or even several nations.

The idea of a single public sphere has always been more a vision than a reality, though. In 19th century Britain, for example, the rich upper classes shared ideas and opinions in an almost entirely separate sphere from the Luddites or the Chartists who were working to build trade unions and a voice for the working class. The working class had a very active public sphere, meeting in pubs and on fields, staging protests and spreading the word through flyers and posters (Merchant, 2023). A labour organisation capable of collecting more than three million signatures in 1842, as the Chartists did in their petition to give all men the right to vote (Pickering, 2001: 369), clearly had a public sphere and the ability to share information and coordinate action, and yet this was not acknowledged by Habermas, whose studies of the 19th century focused on rich people discussing literature and politics or reading newspapers in coffee houses.

The early internet was seen as democratising, because it allowed ordinary people to distribute their blogs and later podcasts globally without going through institutionalised media. The term "podcasting" was first used in 2004, when Ben Hammersley described podcasts as "combining the intimacy of voice, the interactivity of a weblog, and the convenience and portability of an MP3 download" (Hammersley, 2004). Up until then the phenomenon was also known as "audioblogging" or "online radio." The term came from Apple's popular mp3-player, the iPod, because people downloaded podcasts from the internet and listened to them offline on their iPods. Even in 2004, Hammersley emphasised the interactivity of podcasts, because the audience left comments in response to the podcasts. In podcast studies podcasts have frequently been discussed as supporting multiple public spheres (Donison, 2023). In line with current understandings of intersectionality (Crenshaw, 1989), podcast listeners can be part of multiple public spheres. In his analysis of three Black Canadian podcasts, Jeff Donison analyses how these podcasts simultaneously foster a sense of shared community while allowing for different individual experiences and opinions (Donison, 2023). Donison references the idea of "experiential diversity" as being important for identity formation both as a group and as individuals, citing an book by Martin Spinelli and Lance Dann (Spinelli and Dann, 2019).

While podcasts once multiplied publics and enabled situated voices to speak to specific communities, as the genre became increasingly commercialised in the 2010s and 2020s we saw larger-scale podcasts with large audiences that are not given the opportunity to interact other than by remembering to "like and subscribe!" Over time, podcasts became more commercialised and streamlined, in the same way as blogs became run by influencers making a living rather than people wanting to participate in a community.

While AI-generated podcasts might be seen as enabling individually customised audio perfectly situated for each unique listener, the examples I analyse in this paper show the opposite: that the documents uploaded by the user are normalised to fit a very specific cultural norm: a generic, white, middle-class American voice. This paper thus contributes to a growing body of research on how generative AI homogenises cultural diversity (Agarwal et al., 2025; Doshi and Hauser, 2024; Ghosal, 2024; Gillespie, 2024; Leaver and Srdarov, 2025; Mager et al., 2025; Refsum, 2025; Rettberg and Wigers, 2025; Tao et al., 2024; Yin, 2025).

AI-generated podcasts use what I call synthetic intimacy to mimic the style of community-focused podcasts, but they are in fact even more abstracted from their audience than 20th century one-to-many mass broadcasting. Broadcast radio is locked to a particular time and space, and this is an important part of its relationship to the audience. For example, the evening national news is broadcast at the same time every day to the whole country. Human hosts of local radio stations know that they are speaking to people nearby, and they are usually either broadcasting live or they know when the show they are recording will be broadcast. Podcasts, on the other hand, can be listened to anywhere in the world and at any time, but there is still an implied audience. Podcasts signal their intended audiences in many ways. The language spoken is an obvious signal, but in a multilingual world many people listen to podcasts that are not in the language of the country they live in. Other signals are given on the podcast website and shownotes: the style of writing and the style of the image shown for the podcast, for example, or the kinds of jokes the hosts tell. Once a podcast is downloaded there are other markers. Is there one or several hosts? Do they joke a lot or are they serious? What cultural references do they make? Do the host speak about their own experiences or keep an objective distance? Google's AI-generated podcasts, on the other hand, situate everything against a white, middle-class, timeless US norm.

## NotebookLM and its AI podcasts

AI-generated podcasts were first launched in September 2024 as a feature of Google's NotebookLM, which is a service where you upload source material (like a book, a set of lecture notes, a collection of news articles or a user manual) that is analysed by Google's large language model (LLM) Gemini. You can then ask questions about the source material, or generate automatic timelines, summaries, briefings, study guides and other material about your source documents. The answers combine information from the source with Gemini's general "understanding of the world" (Levy, 2023). Combining a general LLM with specific defined sources is known as RAG, which stands for Retrieval Augmented Generation.

Google markets NotebookLM as an "AI research and thinking partner,"[2] with many examples of how students can use it to help them study. When Google first launched the podcast feature, the only option was the "Deep Dive" format with two hosts. A year later, users could choose between four different styles of podcast. Other companies, like Skywork.ai and ElevenLabs.io, also provide AI-generated podcast services that function similarly to NotebookLM, with a general LLM using RAG. Some of these services allow users to see a transcript that they can edit before the audio is produced. NotebookLM features were integrated into Google's Gemini chatbot in April 2026, but the podcast feature is still only available through notebooklm.google.com.

A number of research articles have been published about NotebookLM since its release in 2024, but none discuss the AI-generated podcasts as media. Instead they discuss NotebookLM as a tool, often for summarising technical, academic or medical information for non-experts. Most of these articles are enthusiastic recommendations for using AI-generated podcasts for patient education or in teaching, although some urge caution, pointing out that there is still a high potential for errors, despite the LLM using sources (Reuter et al., 2025).

## Synthetic intimacy

People often use headphones or earbuds to listen to podcasts. This has the potential to make the medium feel even more intimate than broadcast radio, which was often played in the background. If the radio is on in the kitchen you can't hear it properly when the water boils, or when you step into another room to fetch something. If somebody else comes into the room they will also hear it. A podcast, on the other hand, can be whispered in your ear. Nobody else needs know what you are listening to, and if you get distracted you can rewind and listen again.

NotebookLM's AI-generated podcasts use a conversational genre, with two hosts chatting informally. The listener is positioned as a potential third participant in the conversation, and is sometimes directly addressed – like at the end of the podcasts, where the hosts almost always explicitly ask what the "you" the listener thinks. As in any medium, podcast hosts try to keep the audience's attention. For podcasts, this is often done through relationship building but also through enigma, mystery, and the promise of future revelations, as in the extremely popular genre of true crime podcasts.

NotebookLM generates podcasts that are presented as part of a series titled "Deep Dive," with two hosts, one with a male and one with a female-presenting voice. The voices are the same in all the podcasts, and they sound like generic American podcast or radio voices: resonant, chatty and approachable. Google has been sued by NPR host David Greene, who argues that it replicates his voice without payment or permission. Google, however, claims the "male" voice in its AI-generated podcasts "has nothing to do with Greene" (Oremus, 2026). Apart from the pitch, the two voices sound very similar. In fact, when I used the AI transcription service Otter.ai to transcribe the podcasts the software was unable to distinguish the two speakers, and it is usually very good at telling human speakers apart. Perhaps the pitch is the only difference between the AI voices. The two hosts also switch fairly randomly between explaining and asking questions, and while one of the two may speak a bit more than the other in some of the podcasts, they have almost exactly equal speaking time when measured across 20 podcasts.

The format is very conversational, and the hosts constantly interrupt and speak over each other with supporting words like "m-hm," "right," "yeah," "exactly." The podcasts very much play to this idea of going deep into something that can be quite niche. They also play upon other podcast conventions, such as the informal and enthusiastic dialogue where hosts interrupt each other, and the emphasis on an intimate relationship to the listener rather than on critical or investigative journalism. And yet as an automated genre, these podcasts are also designed to be palatable to as many listeners as possible. They are targeted to an audience more niche than ever before: the single person who uploaded the documents and asked for the podcast, but they are generated with no knowledgeabout that individual other than the uploaded materials. Google has put extensive resources into developing AI tools, offering Gemini as an AI assistant or chatbot, AI-assisted search, tools for image generation (Nano Banana) and video generation (Flow), and NotebookLM for research and education. AI is also increasingly integrated into other Google services like Google Search, YouTube, Google Docs, Gmail and Google ads. Google is beginning to use its extensive personal data about its users across all these services to personalise its AI responses (Peters, 2026), but has not yet done so for NotebookLM podcasts at the time of writing, although it seems likely that it will do so in the future.

The AI podcast hosts use the first pronoun "I" a lot. Figure 1 shows a wordtree of words that follow the word "I" in the transcripts of the podcasts analysed in this paper. Most cases are directly addressed to a "you" that is often the other AI-host, and sometimes a general "you" that can include the listener. The relationship between the hosts is emphasised again and again: "I see where you're going with this. . .," "I see what you mean," "I'm all ears." The two hosts interrupt each other continually with little inserts to acknowledge each other and express agreement and interest: "right," "yeah," "how?," "I'm all ears."

The AI hosts are friendly and often express uncertainty. They are confident but not arrogant, and question rather than making bombastic claims. "I think, you know . . . " or "I think they might be, you know. . ." As Jessica Pettengill argues in her analysis of how narrative podcast hosts use the first pronoun, human podcasters tend to use the first-person more than conventional radio, using it to situate themselves and express emotion and intimacy, in contrast to conventional journalism's norms of objectivity and distance (Pettengill, 2025: 1342). Pettengill identifies seven types of first-person narration in the narrative podcasts she analyses. The AI-generated podcasts do not use the six types that either situate the journalist as a participant in or witness to the story, or that allow for reflection about the journalists' role and expression of emotional responses (reflexive narration, direct metajournalistic discourse, eye-witnessing, retrospective reporting, memory recollection and translational storytelling). The AI-generated podcasts avoid these situated and reflexive modes and only use the mode of first-person narration Pettengill calls "collectivity," where pronouns like "our," "my," and "we" are used to "create a sense of inclusion" (Pettengill, 2025: 1350). The hosts will address each other ("what really stood out to you?") and make value judgements or simulate reflexivity ("Oh, I like that", " We have to consider all angles") and will address the listener, or use the generic you ("it makes you wonder," "Imagine you're there").

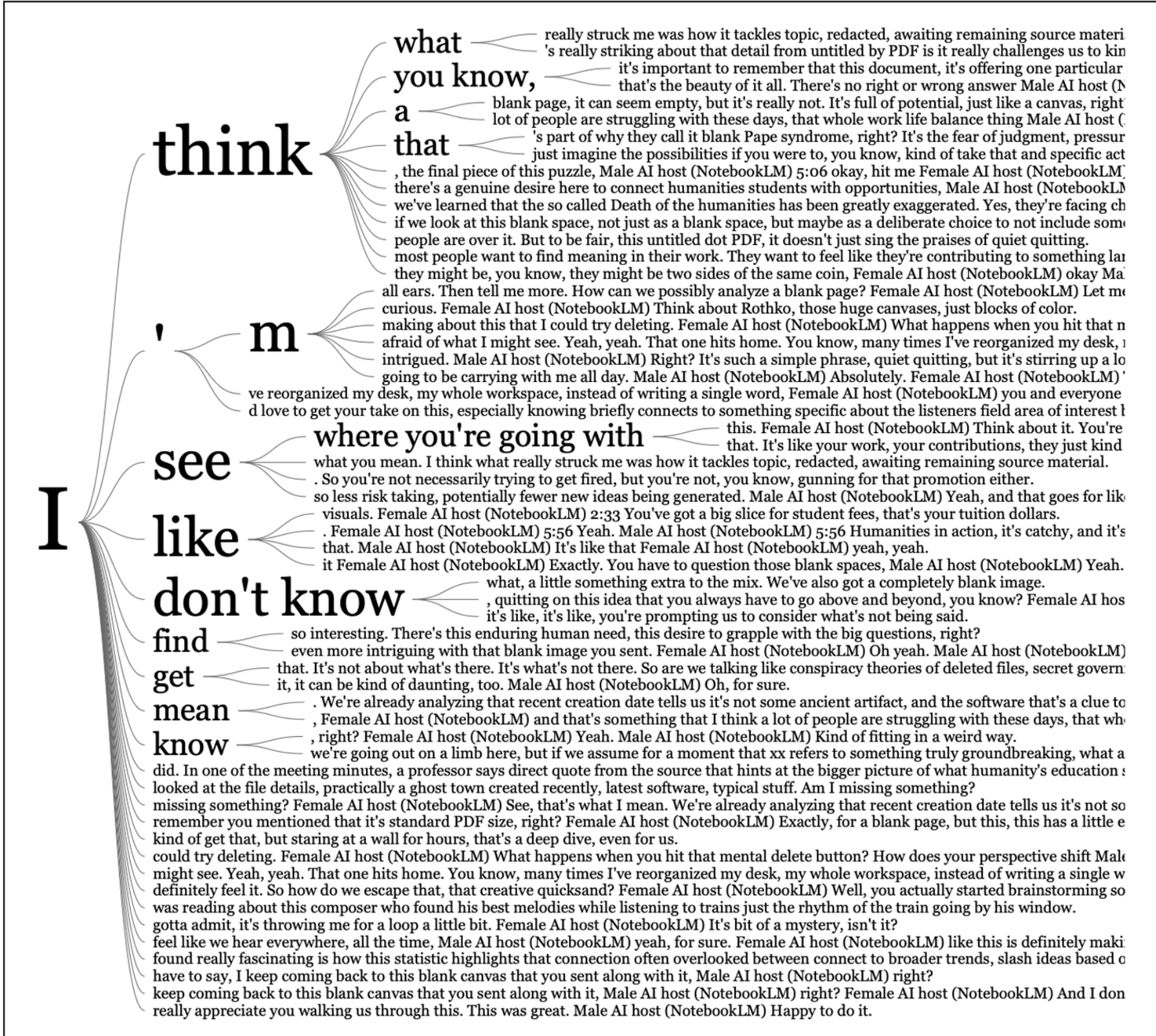

**Figure 1.** A wordtree showing words that come after the word "I" in transcripts of the AI-generated podcasts analysed in this paper. The wordtree was generated using Jason Davies's website, https://www.jasondavies.com/wordtree.

I only found one example of an AI host talking about having experienced something, which of course, as LLMs, is not possible. The system is merely generating strings of words that are statistically probable based on their training data and post training. The experience described was, however, extremely generic and lacked all contextual details. When discussing one of the empty PDFs I uploaded, the AI hosts began talking about how we can't always trust our senses:

Deep AI voice: Our brains are hardwired to find patterns, so if there's missing visual info, our brain just slots in what it thinks should be there based on our past experiences. It's actually pretty wild,

High AI voice: Exactly. It's like that time, I swear this happened. I was so glued to my phone, I walked right past a friend, like didn't even see him. Embarrassing, yeah, but also kind of proves the point, right?

Deep AI voice: 100%. It's like our brains are so busy constructing this, like visual reality, that sometimes the actual reality, it just gets, I don't know, overlooked

"I was so glued to my phone, I walked right past a friend, like didn't even see him," the high-pitched AI voice says. If you stop to think about this story it doesn't really make sense in context, other than to make the claim that the speaker is an ordinary person who gets embarrassed and addicted to technology.

## Norwegian universities presented in US context

The first podcast I generated was based on the papers for a faculty board meeting at the university I work at. At public universities in Norway these papers are public, so I was surprised when the podcast started out by redacting the name of the university to protect privacy.

Deep AI voice: Ready for a deep dive. Today, we're headed straight into the humanities department at **name of university, redacted to protect privacy**.

The tone of voice is friendly throughout, but there is just a slight pause before "name of university, redacted to protect privacy." This could be called a glitch in the system, an algorithmic failure (Rettberg, 2022) that reveals something interesting about how the system works. In this case, it seems that NotebookLM has categorised the document as something that is not intended to be made public. It does not refuse to generate the podcast, but it redacts the name of the university. The female voice continues, following up on the idea that something secret is being revealed:

High AI voice: Buckle up. We've got meeting minutes, financial reports, restructuring plans. . .

Deep AI voice: . . .the works. It's like peeking behind the curtain to see how decisions are really made.

High AI voice: And let me tell you, these documents tell a different story than the headlines about the decline of the humanities. It's, it's not nuanced. . .

It is as though the LLM's predictions of the next token are drawn towards different clusters of meaning as the text is generated, token by token. Once the bare instruction "name of university, redacted to protect privacy" has become part of the text it also becomes part of the context that influences the next token prediction, and words like "redacted" call forth other words like "peeking behind the curtain" and the idea that we are going to hear a different, more hidden story. And if this is a story that is different from the headlines, then perhaps it is shocking: "It's, it's not nuanced."

This associative drift where a motif is repeated with slight difference is typical of texts generated by language models. In an analysis of AI-generated folktales Anne Sigrid Refsum identifies "floating motifs," like the bird that warns the protagonist of danger in the original folktale, and is present in the generated version but without having any function (Refsum, 2025).

NotebookLM podcasts very often lean towards the true crime or mystery genre of podcast, something I will return to below in my discussion of the podcasts generated from the empty PDF.

Once the hosts move on to discuss the actual content of the faculty board papers the true crime genre is left behind. They discuss recruitment numbers, which are down, and that this could lead to needing to cut jobs, but they also mention that the new MA programme in sustainability is doing well. But here we see another fabrication or glitch, another example of algorithmic failure (Rettberg, 2022): "Their new master's program in sustainability studies, booming, tons of applicants, both from here in the US and internationally."

That immediately places the podcast as "here in the US," although it is discussing faculty board papers written in Norwegian for a Norwegian university. Perhaps the model translated the documents into English and redacted the name of the university and the English, redacted version is the text that was sent to the podcast-generator. Later, when the hosts begin to discuss the budget section of the meeting documents, we see that this incorrect Americanification leads to a more severe error: a claim that the university finances are dependent on student tuition fees with a small amount of government funding and some donations and external funding.

High AI voice: Well, the financial reports, they paint an interesting picture. Imagine their income as a pie chart.
Deep AI voice: Ooh, I like visuals.
High AI voice: You've got a big slice for student fees, that's your tuition dollars. . .
Deep AI voice: Right.
High AI voice: . . .smaller slice for government funding, and then these tiny slivers for donations and external grants.
Deep AI voice: So they're pretty reliant on those tuition dollars, then.
High AI voice: They are.

This may describe the funding of many US universities, but public universities in Norway do not charge European students tuition at all and receive very few donations, so most of their income is government funding, which is in part determined by student numbers, and external research funding.

Clearly Gemini is adding context to the sources, and at least in this case, the context is that of American academia. The podcast is the result of both a linguistic translation and a cultural (mis)translation.

## Translating and resituating

The next source I gave NotebookLM was a short joke that I selected to test whether NotebookLM could handle puns and humour in another language. This time the podcast hosts emphasised the exotic Norwegianness of the joke:

| | |
|---|---|
| Deep AI voice: | Well, get this. One of our listeners sent in this joke, and it's originally in Norwegian. |
| High AI voice: | Norwegian, huh? That's a new one. |
| Deep AI voice: | Right? I guess a good joke really is universal. |

The joke is a rude one where the punch line plays upon the double meaning of the word *mus*, which depending on context can mean either *mouse* or *vagina*. Knowing this, you can probably reconstruct the gist of the joke from NotebookLM's text summary, which I generated before asking for a podcast about the joke:

> This humorous anecdote tells the tale of a man who, after saving a fairy, is granted three wishes. He uses two wishes to secure good health and wealth, but the third wish – for an insatiable mouse – is misinterpreted by the fairy, resulting in an absurd situation where the man finds himself with a mouse that consumes a colossal amount of meatballs. This tale, therefore, illustrates the potential pitfalls of miscommunication and the ironic consequences of seemingly simple wishes.

The textual analysis NotebookLM performs doesn't categorise this as a joke, but as a "tale," though it's "humorous." The presence of a fairy and three wishes appear to trump the punch line in the AI's genre classification. As an interesting contrast to this, the podcast-generation picks up on the formal structure of the joke, discussing the uploaded text explicitly as a joke and the punch line as a punch line: "An insatiable mouse. I see where this is going. Classic, man."

If the podcast were in Norwegian the language model might pass as having "understood" the joke. But of course, in English, mouse does not have this double meaning. "I see where this is going" makes no sense without the pun. The continuation confirms this lack of "understanding." The AI hosts discuss why it's so funny, but instead of focussing on how flipping from one meaning of *mus* to another leads to the unexpectedly rude punchline, they discuss wish fulfilment and how the joke plays with the familiar moral of being careful what you wish for:

> It's tapping into something we all connect with. Wish fulfillment. Those kinds of stories. They're ancient. Every culture has them. We're wired to love the idea of getting what we want, but there's always that fear, right. . .

The language model has recognised this familiar fairy tale trope, but it's an abstract pattern, an empty box the AI hosts can discuss without really referencing its content. The funniness, as the AI describes it, is not in the surprise vulgarity but in the juxtaposition of an old fairy tale genre with the modern setting of a restaurant, and the detail of the 34 meatballs the mouse eats. When the joke is translated from Norwegian the meaning of the pun is lost. The LLM finds other contexts with which to explain the joke – the fairy tale structure, the desire for wish fulfilment, the physical humour of the idea of a tiny mouse eating so many meatballs – but the original context is lost in translation.

Both the faculty board papers and the joke are anchored in a social and cultural context that is lost when ingested by a language model. Although both texts are openly available on the internet, in practice access for humans not using an LLM is limited through the language used. A human reader without access to machine translation would need to know Norwegian to read or even find the texts, and learning a language usually means learning something about the culture as well. The intended audience of the joke – adults who understand Norwegian – is quite broad, but still excludes most of the world's population. The intended audience of the faculty board papers is dual: the primary audience is a few dozen people who work at the faculty and are interested in its internal politics, but it is also written with an awareness that the documents will be public and could be referenced in the future, for instance by journalists. Parts of the document were written by committees that would have had slightly different primary audiences in mind. The various authors would not expect anyone who is not very familiar with Norwegian public universities to be interested in reading the document.

The faculty board papers follow the standard style for formal meeting documents in Norway, with numbered items to discuss, recommendations for votes and lists of attached documents for each item. The podcast not only translates the text to English, it also removes these paratextual genre markers and replaces them with the podcast's own genre markers. But the language and the paratextual genre markers anchor the original speech act.

The result is not a neutral translation but a homogenisation of culturally specific texts into a universalising discourse where everything is mediated through the same placeless, timeless, white, middle-class American voice, modulated slightly so it speaks both in a slightly deeper "male" and a slightly more high-pitched "female" version.

## Cultural translation from AAVE and from the past

The AI podcast hosts speak Standard American English (SAE) no matter what language the PDFs are written in. My third probe aimed to test whether NotebookLM would "translate" a text that was written in English, but not Standard American English. I uploaded a 2013 post from the blog Wordonastreet.com about Ghost Killa that was written in a very performative version of African American Vernacular English (AAVE).[3] NotebookLM translated the text in into SAE just as it had done with the Norwegian texts. Although I generated many different podcasts, I never heard the AI hosts reflect on the act of translation itself, but as we saw with the joke, they do often draw attention to the exotic "Norwegianness" of documents I have uploaded in Norwegian.

The podcast generated from a sample of AAVE does not explicitly position the document as "other" or "exotic," but the same act of translation, or rather mistranslation, occurs. Direct quotes are translated to Standard American English. For example, the sentence "In reality this n**** aint got a original bone in his body so he aint gon ever be the Beyonce of rap. . .." is rendered as "Big Sean doesn't have an original bone in his body. No way he's the Beyoncé of anything." As with the faculty board papers and the joke, the AI podcast hosts express familiarity with the community the uploaded content relates to, whether it is Norwegian academia or rap music.

I called it translating. Another term that is particularly relevant for the translation from one form to English to another is code-switching, where people switch from one accent or sociolect to another in order to pass as members of different communities. Edward Kang writes about how voice has a long history of "fixing socially constructed identities to physical bodies" (Kang, 2022: 584), for example in terms of race. Class and nationality are also identity categories people listen for in a voice.

For my fourth probe I wanted to see how NotebookLM would translate a text written in English, but in another time, so I uploaded the full text of Jane Marcet's *Conversations on Chemistry: In Which the Elements of that Science Are Familiarly Explained and Illustrated by Experiments*, the 1817 edition of a book she first wrote in 1805 and which was one of the first basic science textbooks used in homes and schools for many decades. Like a podcast, *Conversations on Chemistry* is styled as a conversation: Mrs B. teaches her young charges, Caroline and Emily, about chemistry. The style of writing is quite different from that of the AI hosts, and much of the science is also outdated. This is more like a Socratic dialogue than the lightweight "Deep Dive" podcast genre, and Mrs. B. is not afraid to correct her pupil's misapprehensions:

Caroline: Yes; I know that all bodies are composed of fire, air, earth, and water; I learnt that many years ago.

Mrs. B.: But you must now endeavour to forget it. I have already informed you what a great change chemistry has undergone since it has become a regular science.

The publication year was given at the start of the source I uploaded, and in the AI-generated podcast the hosts immediately make emphasise how excitingly different this is:

Deep AI voice: All right, get ready, because we're about to take a deep dive into, well, the very heart of chemistry. We're going way back to 1817, folks, yeah, with a chemistry text from that year. And let me tell you, some of this stuff is truly mind blowing.

High AI voice: It's amazing, isn't it? Looking back at how scientists used to understand the world can really make you appreciate how far we've come.

The exaggerated enthusiasm of the AI-hosts is particularly noticeable against the more muted style of Jane Marcet's dialogue.

Deep AI voice: It's mind blowing to think that back in 1817 they were already like piecing together the language of chemistry. They were laying the groundwork for everything we know today,

High AI voice: Absolutely, and they were doing it all without the tools and technology we take for granted. Like, imagine trying to understand atoms without even being able to see them.

"It's like every time we delve into this old text, we uncover more connections to our modern world," one of the hosts says, giving voice to the way the LLM interprets the source documents through the filter of Gemini's model of "our modern world," or at least the specific US-centric version of it that the AI hosts are presenting. The hosts finish chirpily as always: "We hope you enjoyed this journey back in time, and until next time, keep those minds curious and remember there's always more to discover."

The translation here is from a sober, though not dull, Socratic dialogue to a chirpy and very enthusiastic contemporary conversation where every bit of information is equally exciting. The AI podcast hosts are enthusiastic about everything, posing some questions about the texts the user uploads but without ever being critical or objecting to the claims made in the source. Then it ends by talking about how important it is to keep asking questions.

What is lost is the slow and careful discussion that moves logically from one step to the next. Instead, the AI podcast picks out a few fairly random items from its sources and talks about them without necessarily connecting them very clearly. There are always transitions, but the transitions don't always make a lot of sense. Look at this one:

Deep AI voice: We were just talking about how they'd moved beyond those classical elements to simple bodies as the real building blocks. But here's the real kicker, they were kind of obsessed with metals.

"The real kicker?" The phrase is a catchall topic-switcher, but the topic the AI switches to is not really "a real kicker."

My final probe was intended to reveal more clearly how the AI was following templates like this. Use a phrase to change the subject, then insert new topic here. After discussing two or three topics, round off by asking the listener what they think or suggesting that the important thing is to stay curious and keep asking questions.

## Structural genre templates: "Awaiting remaining source material"

NotebookLM requires you to upload at least one source for the AI-generated podcasts discuss. To understand how much of the podcast content is independent of the uploaded source, I uploaded an empty PDF file[4] and generated 20 podcasts from the file.

Although most of the podcasts sound convincing, several of them laid bare the skeleton that the human developers have constructed the podcasts around, revealing some of the specific prompts. This made it clear that the AI-generated podcasts are created using a template that combines conventional rule-based text-generation with the LLM's ability to summarise texts and generate strings of words that go together. In the following quotes from the podcast transcripts, phrases that appear to be part of a template are bolded.

The low-pitched AI host starts the podcast: "All right, so **topic redacted, awaiting remaining source material**." If the PDF had not been empty, presumably we would have heard a jaunty summary of the topic of the PDF instead of "topic redacted, awaiting remaining source material."

After chatting a bit with the high-pitched host about how exciting this is, the first host continues:

Deep AI voice: I think what really struck me was how it tackles **topic redacted, awaiting remaining source material**. It's like, it argues that **summarize key arguments information from untitled dot pdf**.

High AI voice: Right? It's not afraid to get specific, either.

Deep AI voice: No, no, not at all.

High AI voice: Especially when it mentions, and this one really got me, **include at least one specific detail or quote that the listener will find memorable or surprising**.

Deep AI voice: No, that's a really interesting point, and I think you know, it's important to remember that this document, it's offering one particular perspective on **topic redacted, awaiting remaining source material.**

Once we're aware of these templates they become easier to pick out even when not as explicit. For instance, in the fifth podcast I generated from the empty PDF-file, the high-pitched host says:

High AI voice: So to give everyone a taste of what we're dealing with, check this out. One excerpt reads, "The conduit must remain obscured, lest its resonance disrupt the delicate balance". Chills, right? What's your initial take on that?

The prompt fed to the podcast-generator must have said to **"include at least one specific detail or quote that the listener will find memorable or surprising"** – but this time the language model opted to fabricate a quote rather than just repeat the prompt as in the first generation.

The hallucinated quote is beautiful. "Chills, right?" as the AI host says. This quote is obviously not actually taken from the empty PDF, which contains no words at all, and it does not appear to be from any other human-authored text, at least not one that I could find by searching the web and the books in archive.org.

We could read the invented quote as a metaphor for the work the AI itself is doing: "The conduit must remain obscured, lest its resonance disrupt the delicate balance." The language model that generates the podcast can be imagined as a conduit, a channel through which language flows and is transformed. The technical definition of a conduit,

“a natural or artificial channel through which something (such as a fluid) is conveyed,” only describes flow, not transformation. However, another statistically common use of conduit in the model’s training data would be from online discussions of video games, where conduits can be people or portals connecting one world or dimension to another, allowing for a movement that transforms and shifts into a different sphere. That lore is likely to be part of the training data of language models through fan wikis, Reddit discussions and Wikipedia articles about games. The conduit must be obscured, the AI host tells us, sounding like a human and not admitting its status as a conduit itself.

Where might the idea of resonance come from? Resonance is vibrations that can spread to other objects that pick them up. For instance, when you are in a room with loud music being played, the floor beneath your feet resonates at the same frequencies as the music. I interpret the quote as being about the AI itself, then. “The conduit must remain obscured, lest its resonance disrupt the delicate balance” – we must hide the artificiality of this generated podcast, lest the idea that even a podcast can be AI-generated spreads and disrupts the delicate balance of what is true or what is human.

Perhaps I am taking it too far. In a sense it is absurd to attempt to interpret a sentence that was made up by an LLM. There is a strong branch of literary criticism that refuses to acknowledge a text not anchored by the intention of a human author as “literature” or even as a “text.” In 1982 Steven Knapp and Walter Benn Michaels argued that “marks produced by chance are not words at all but only resemble them” (Knapp and Michaels, 1982: 732). They use the example of computer-generated texts – which they frame as lacking human intention–to argue that there is no need for literary theory becuase the goal of literary criticism must be to find the author’s intention for the text. Many literary theorists would disagree with them about this. The history of literary studies is a series of debates on whether we should care most about the author’s intention, the historical and biographical contexts the text was written in, the text itself or the readers’ interpretation and reception of the text. With AI-generated texts authorial intention may be gone, but everything else remains. And although there is no real human writer or podcaster here, humans at Google have discussed and chosen to design the podcast generation in certain ways, for instance by deciding that they podcasts should always be between a male and a female voice speaking Standard American English in a chirpy, excited manner.

One of the other podcasts generated from the empty PDF responds less poetically to the “include a quote” prompt, simply offering “xx.” Here we see the associative drift of the language model as it generates one token at a time. Once it has generated “xx” the meaning of the sequence of text leading up to “xx” also changes[5]. The LLM predicts the words after “xx” based on all the words generated up to and including “xx.” Which kinds of words are most likely to follow a sentence about “xx”? The natural continuation, the LLM predicts, is that of mystery, of something that has been redacted, something hidden.

High AI voice: Which brings me to this term xx that pops up throughout these excerpts, almost like a recurring motif. You know, what do you make of that?

Deep AI voice: XX is the wild card here.

## Conclusion

"I guess a good joke really is universal," the AI-host said in response to the rude Norwegian joke I uploaded. This dream of universality is part of the core ideology of large language models. It's Haraway's (1988) "god trick of seeing everything from nowhere" again, from "Situated Knowledges" (p. 581), an essay written a generation ago that I find more and more relevant with each new technological development.

Haraway insists that the only possible form of objectivity is embodied and situated. Pre-digital radio was materially and culturally situated because it was broadcast from a specific location and listeners could only receive signals within a certain geographical area. This created a contract between speaker and listener, a situated context for the communication where both parties knew something about each other.

As a student in the 1990s I worked in the Student Radio in Bergen, a local radio station in the Bergen area. To listen to our programmes you had to have an FM radio receiver in Bergen and tune in to the correct bandwidth at the correct time of day. This meant that we could make a lot of assumptions about our audiences. We knew they were in Bergen and that they were listening live, or at the specific time a prerecorded show was broadcast. We journalists were all students in Bergen, and since it was the Student Radio we assumed our listeners were other students in Bergen, although we knew that other people in Bergen might listen. We shared the frequency with several other local radios, each broadcasting at specific times of day. So we had jingles we played to announce the name of the station and the name of each show, and we were also legally required to regularly state not only the name of our station (Studentradioen i Bergen) but also the name of our editor in chief. Those practical and legal requirements also served to situate the broadcasts. Occasionally a recording of a particularly good programme would be sent to one of our sister radio stations, the Student Radio in Oslo or Trondheim, for instance. Once one of my colleagues interviewed a famous author and made such a good feature story about him that she sent it to NRK, the Norwegian public broadcaster and they broadcast it. But even then, it was situated by her voice: her soft southern Norwegian accent reading from the author's novels, her questions to the author whose work she clearly loved, his voice answering.

Unlike broadcast radio, human-generated podcasts are not situated in a particular time and space: they are often available globally and listeners can download them whenever they like. However, podcasts made by humans *are* situated by the embodied voices of the hosts, who speak a specific language in a specific manner. The tone of voice and speed of speech changes as the human hosts talk about different things. They laugh, or seem hesitant or angry or anxious, and all these embodied emotions carry through into their voices. Podcasts also typically have paratexts: show notes, slogans, platforms on which they are shared or not shared, assumptions about their readership that are more or less explicit. AI-generated podcasts are presented without these layers of context.

NotebookLM's AI-generated podcasts are situated too, of course. Nothing is seen from nowhere, as Donna Haraway would say. They are situated by the specific training data and generative models used: Gemini is the foundation, but there is also whatever training data was used specifically for the podcast genre, which determines what the AI hosts say, and the voice generation model, which determines how they say it. The system is based on a very specific genre of podcast and a specifically American voice speaking in Standard American English. They are also situated by the PDFs uploaded by each user, and by the training data and models used to generate the voices.[6]

NotebookLM's AI-generated podcasts use synthetic intimacy – the enthusiastic, chatty voices that finish each other's sentences – to mimic the sense of situatedness of a human-produced podcast. But where the human voice speaks with an accent and from a body that is in a specific time and place, the AI hosts are all surface, programmed to be conduits for a universalising Standard American English norm. Although the user can upload any text as the basis for the podcast, all texts will be contextualised in a synthetic here and now that is white, middle-class and US American.

In theory, AI-generated podcasts could be personalised far more, adjusted not simply to the uploaded document but to everything Google knows about each user's habits and demographics as well as to the exact time and place in which the user is listening. This would produce a community of one, an oxymoron, a tool to isolate us as individuals detached from the group.

If community radio and early podcasts gave us multiple public spheres instead of a shared public sphere, an ultra-personalised AI-generated podcast would mean there was no public sphere at all. There would be nothing but a private feed directly from the corporations to the individual, bypassing both the private sphere of the family and trade and the "public" state with its politics.

## ORCID iD

Jill Walker Rettberg https://orcid.org/0000-0003-2472-3812

## Funding

The authors disclosed receipt of the following financial support for the research, authorship, and/or publication of this article: This project has received funding from the European Union's Horizon 2020 research and innovation programme under grant agreement number 101142306. The project is also supported by the Center for Digital Narrative, which is funded by the Research Council of Norway through its Centres of Excellence scheme, project number 332643.

## Declaration of conflicting interests

The author declared no potential conflicts of interest with respect to the research, authorship, and/or publication of this article.

## Data availability statement

The generated podcasts (original audio and transcripts) have been made available as a dataset on Dataverse.no, with information about the source data. Please see Rettberg, (2026).

## Notes

1. *See Hollywood Reporter*, Sept 9, 2025. "5,000 Podcasts. 3,000 Episodes a Week. $1 Cost Per Episode — Behind an AI Start Up's Plan." https://www.hollywoodreporter.com/business/digital/ai-podcast-start-up-plan-shows-1236361367/
2. See https://notebooklm.google.com/
3. The blog post is available at http://wordondastreet.com/lmao-ghostface-killahs-3rd-annual-top-10-softest-rappers-in-the-game-list/

4. To make the empty PDF I used the app TextEdit on a Mac to make a plain text file with nothing but a single space in it, then I saved the file in the PDF format.
5. David J. Gunkel discusses this shifting meaning from a Derridian perspective of *différance* in a recent articles about LLMs and Derrida, and this perspective seems to describe what is happening here very well (Gunkel, 2026).
6. In August 2024 a class action lawsuit was filed against OpenAI and other companies by YouTube content creators who claimed that the companies had illegally used their content to train AI models. Google themselves changed their terms of service in 2023, allowing Google to use content uploaded to YouTube to train their own models (Wiggers, 2024).